\documentclass[twocolumn,aps,pre,floatfix,superscriptaddress]{revtex4-1}
\usepackage[export]{adjustbox} 
\usepackage[toc,page]{appendix}
\usepackage{epsfig}
\usepackage{color}
\usepackage{braket}
\usepackage{graphicx}
\usepackage{booktabs}
\usepackage{subfigure}
\usepackage{blindtext}
\usepackage{amssymb,amsfonts,amsmath}
\usepackage{tikz}

\begin{document}

\title{Build up of yield stress fluids via chaotic emulsification}

\author{Ivan Girotto}

\affiliation{The Abdus Salam, International Centre for Theoretical Physics - Strada Costiera, 11, 34151 Trieste, Italy and Department of Applied Physics, Eindhoven University of Technology - 5600 MB Eindhoven, The Netherlands and University of Modena and Reggio Emilia - Via Universit\'a, 4, 41121 Modena, Italy}

\author{Roberto Benzi}

\affiliation{Department of Physics and INFN, University of Tor Vergata - Via della Ricerca Scientifica 1, I-00133 Rome, Italy}

\author{Gianluca Di Staso}

\affiliation{Department of Applied Physics, Eindhoven University of Technology, 5600 MB Eindhoven, The Netherlands}

\author{Andrea Scagliarini}

\affiliation{CNR-IAC, - Via dei Taurini 19 I-00185 Rome, Italy}

\author{Sebastiano Fabio Schifano}

\affiliation{University of Ferrara and INFN Sezione di Ferrara - via Saragat 1, I-44121 Ferrara, Italy}

\author{Federico Toschi}

\affiliation{Department of Applied Physics, Eindhoven University of Technology - 5600 MB Eindhoven, The Netherlands and CNR-IAC, - Via dei Taurini 19 I-00185 Rome, Italy}

\date{\today}

\begin{abstract}
  \noindent Stabilized dense emulsions display a rich phenomenology
  connecting microstructure and rheology. In this
  work we study how an emulsion with a finite yield stress can be
  built via large-scale stirring. By gradually increasing the volume fraction of the
  dispersed minority phase, under the constant action of a
  stirring force, we are able to achieve volume fraction
  close to $80\%$. Despite the fact that our system is highly
  concentrated and not yet turbulent we observe a droplet size
  distribution consistent with the $-10/3$ scaling, often associated
  to inertial range droplets breakup. We report that the
    polydispersity of droplet sizes correlates with the dynamics of
    the emulsion formation process. Additionally we quantify the
  visco-elastic properties of the dense emulsion finally obtained and
  we demonstrate the presence of a finite yield stress. The approach
  reported can pave the way to a quantitative understanding of the
  complex interplay between the dynamics of mesoscale constituents and the 
  large scale flow properties yield-stress fluids.
\end{abstract}


\maketitle

\section{Introduction}
\noindent Emulsions, the dispersion of (at least) one liquid component into
another in the form of droplets \cite{tadros2013emulsion}, are common to a vast number of
products and applications, including food, cosmetics and oil
industries \cite{gallegos1999rheology,mcclements2015food,chang1999isothermal,egolf2005iceslurries,coussot2005rheometry}. 
The physics of emulsions presents many outstanding
scientific challenges, connected with their rich phenomenology,
which ranges from that of a non-Newtonian viscous fluid to an
elastic solid \cite{larson1998structure} as the result of the strong
coupling between the microscopic and macroscopic physics \cite{barnes1994rheology,pal1996effect,mason1999new,derkach2009rheology}. 
Despite the fact that much of this phenomenology seems
universal and common to many soft-glassy materials (such as foams, gels, slurries, etc), the specific
mechanical response of the emulsion strongly depends on its microstructure \cite{pal1996effect} and
a fully quantitative theoretical understanding of the rheology-microstructure links is still lacking.
Achieving certain rheological properties
relies heavily on the capability to properly control how the emulsion is built.
Nevertheless, much of the existing body of works (both experimental and numerical) 
focused rather on the effect of the physico-chemical properties of the liquids and surfactants,
yielding a certain droplet elasticity or interfacial rheology 
\cite{derkach2009rheology,vankova2007emulsification1,vankova2007emulsification2,vankova2007emulsification3}.
In particular, in all the previous numerical studies of soft-glassy rheology, the system 
is assembled by simple juxtaposition of the elementary meso-constituents, with a given mean size 
and polydispersity, but according to an arbitrary distribution.\\
\begin{figure}[htbp]
\centering
\subfigure[ $\phi=30\%$ $t=0.6 T_L$]{
  \includegraphics[width=0.17\textwidth]{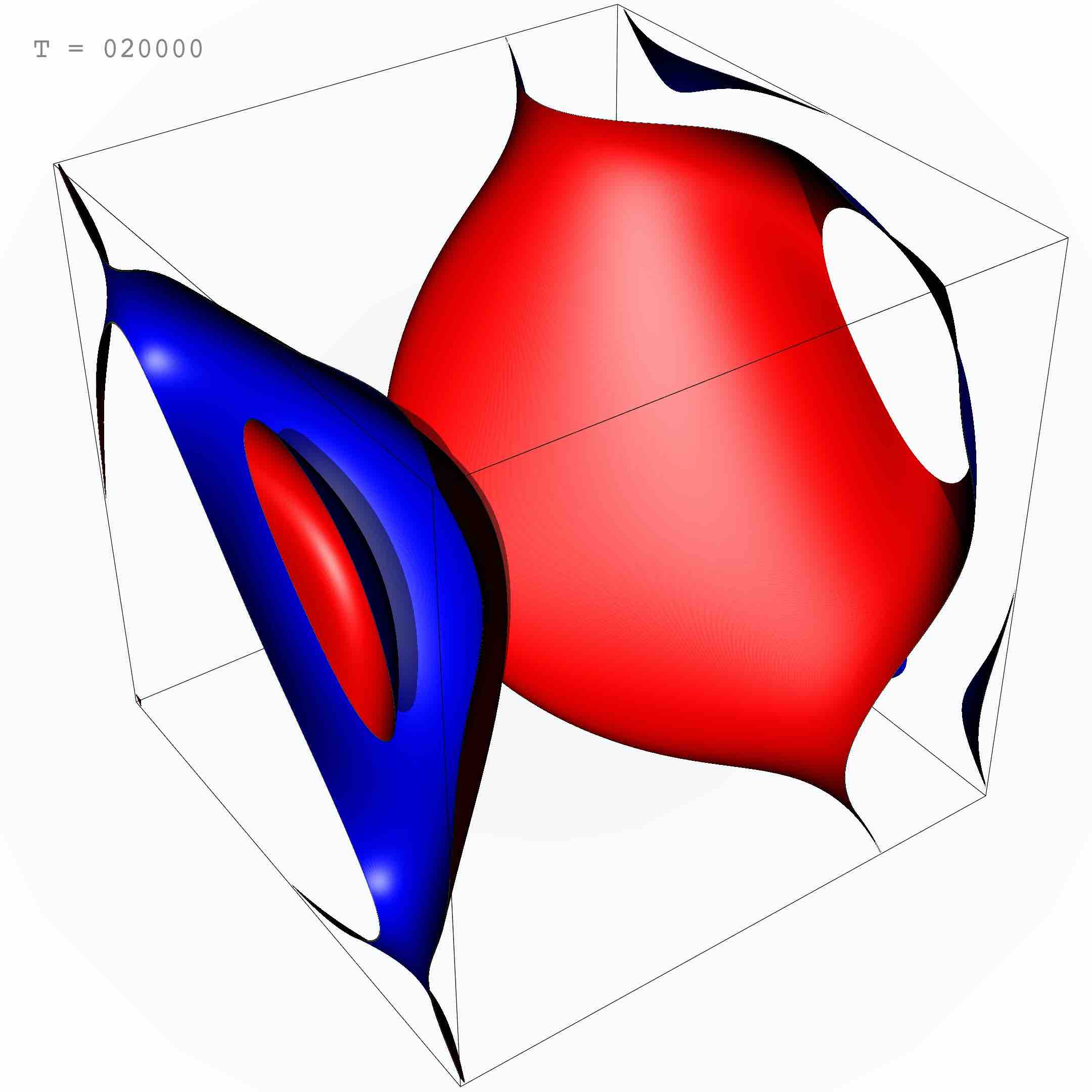}
  \label{figure:motion_a}
}
\subfigure[ $\phi=31\%$ $t=1.2 T_L$]{
  \includegraphics[width=0.17\textwidth]{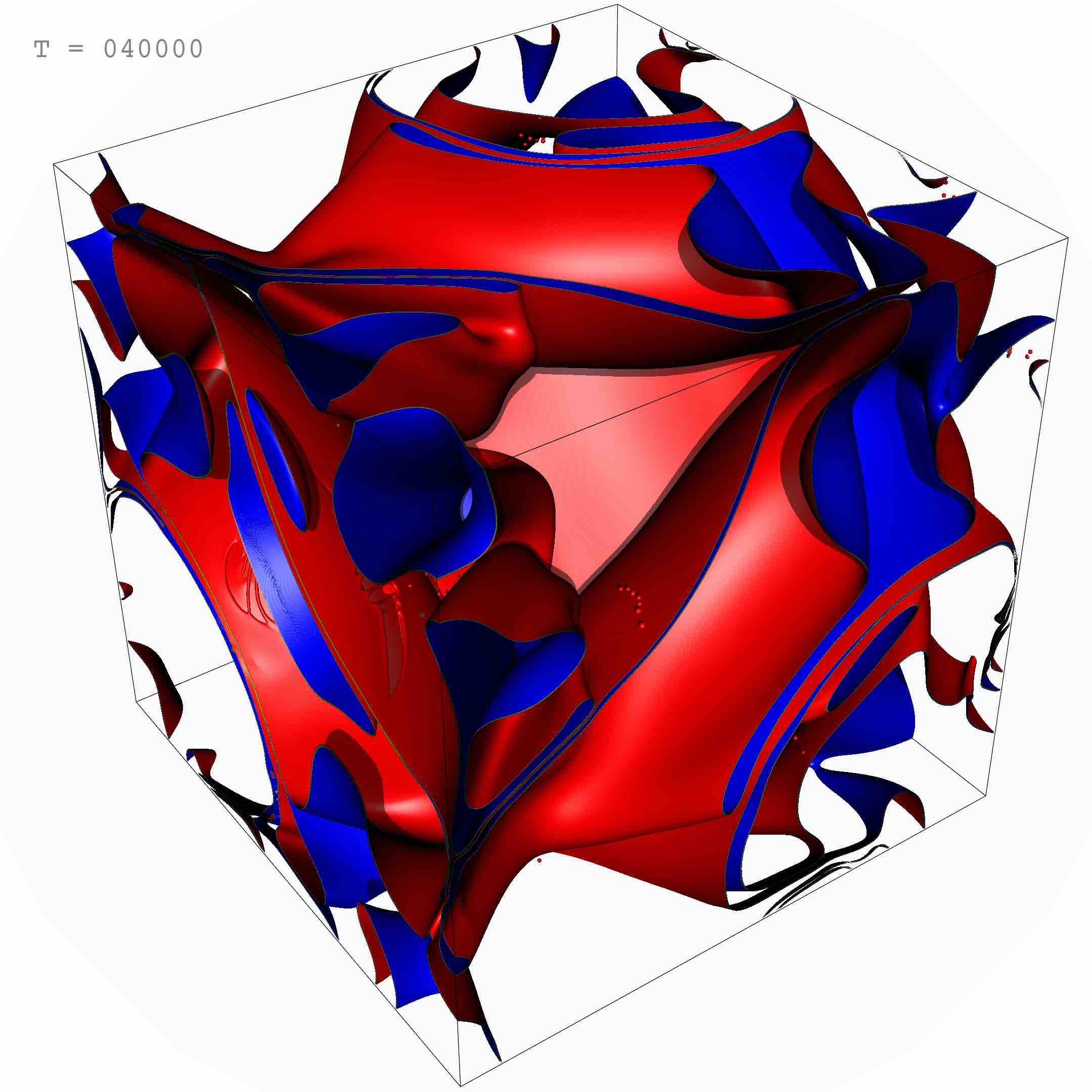}
  \label{figure:motion_b}
}
\subfigure[ $\phi=32\%$ $t=1.8 T_L$]{
 \includegraphics[width=0.17\textwidth]{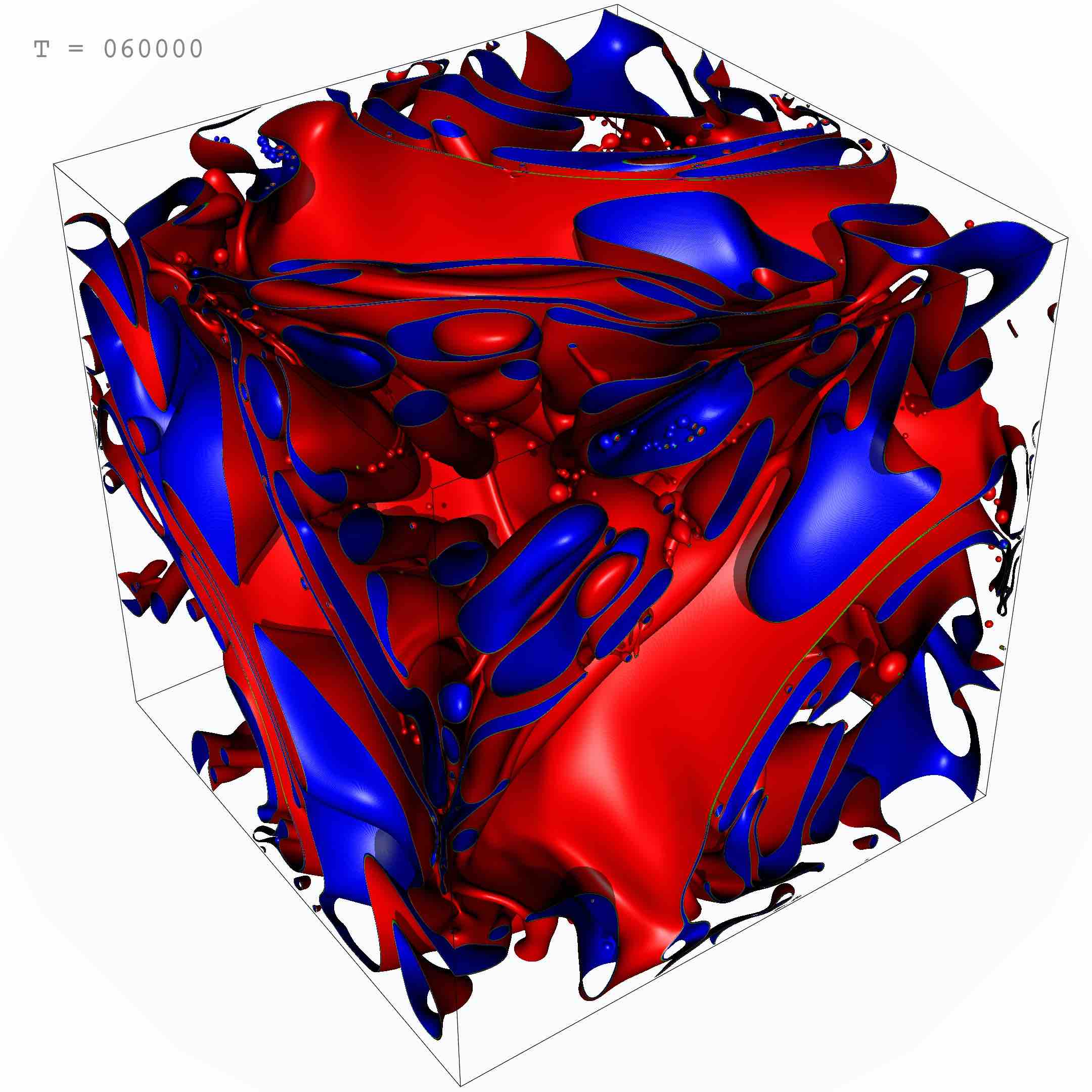}
 \label{figure:motion_c}
}
\subfigure[ $\phi=33\%$ $t=2.4 T_L$]{
 \includegraphics[width=0.17\textwidth]{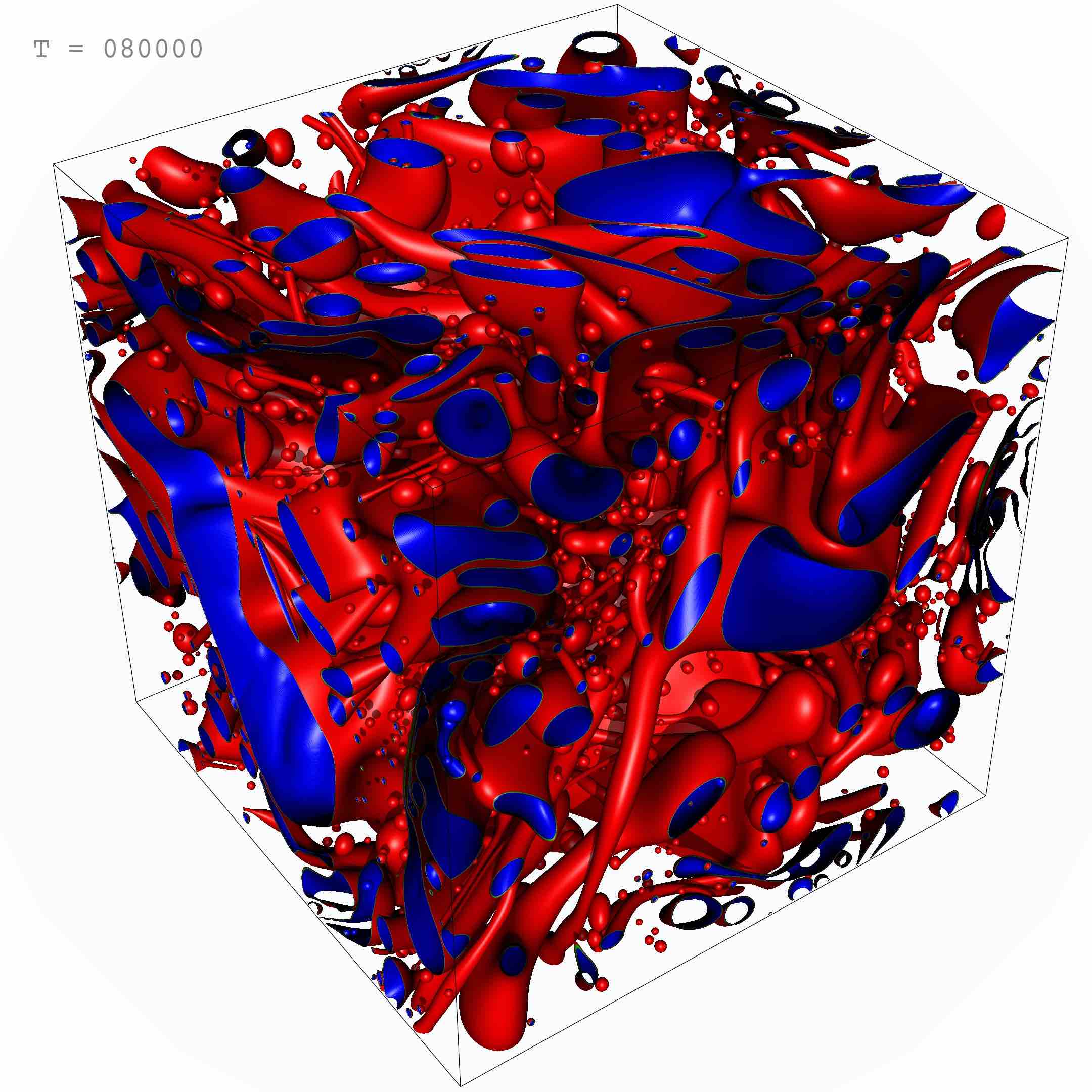}
 \label{figure:motion_d}
}
\subfigure[ $\phi=35\%$ $t=3 T_L$]{
  \includegraphics[width=0.17\textwidth]{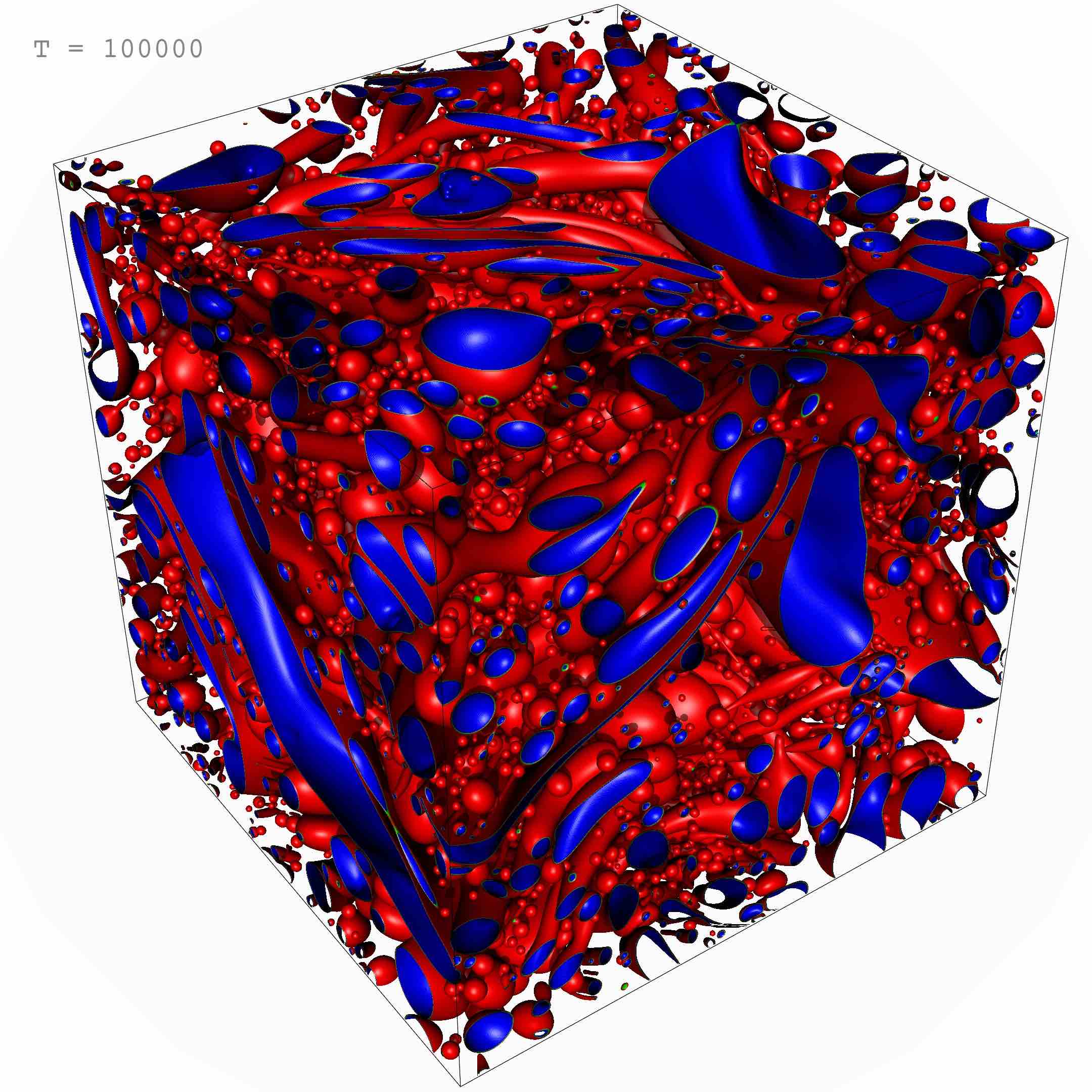}
  \label{figure:motion_e}
}
\subfigure[ $\phi=42\%$ $t=6 T_L$]{
  \includegraphics[width=0.17\textwidth]{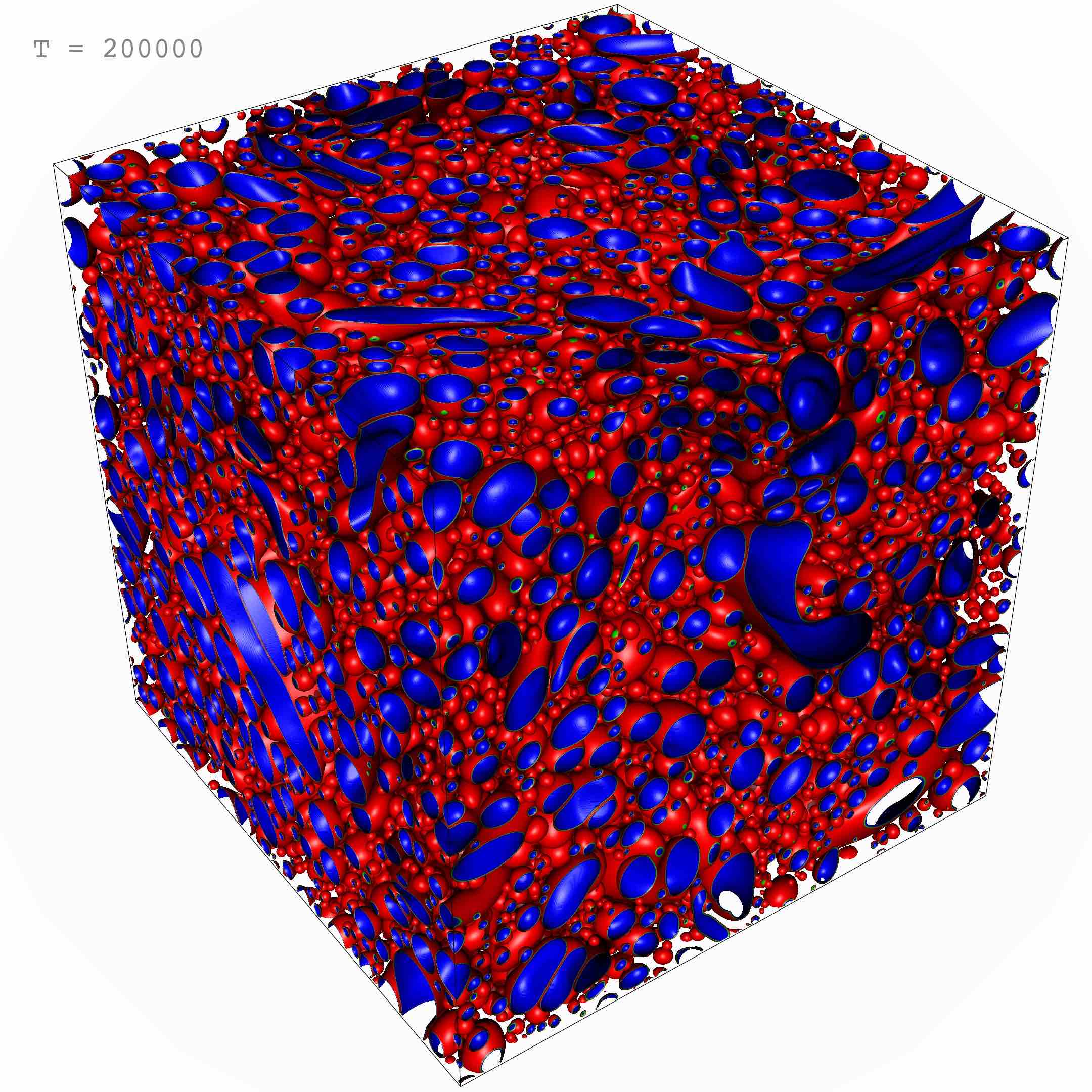}
  \label{figure:motion_f}
}
\subfigure[ $\phi=50\%$ $t=9 T_L$]{
  \includegraphics[width=0.17\textwidth]{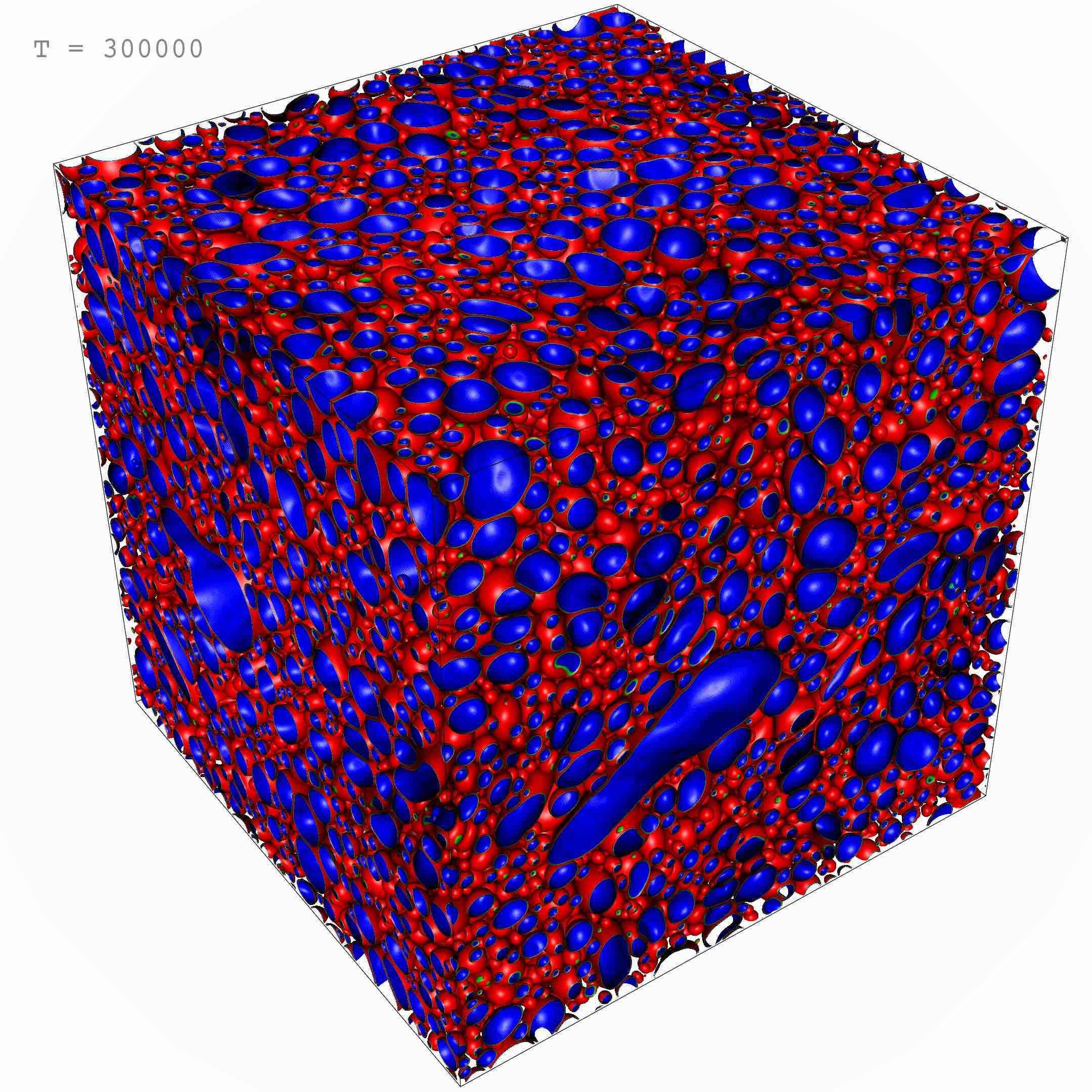}
  \label{figure:motion_g}
}
\subfigure[ $\phi=60\%$ $t=12 T_L$]{
  \includegraphics[width=0.17\textwidth]{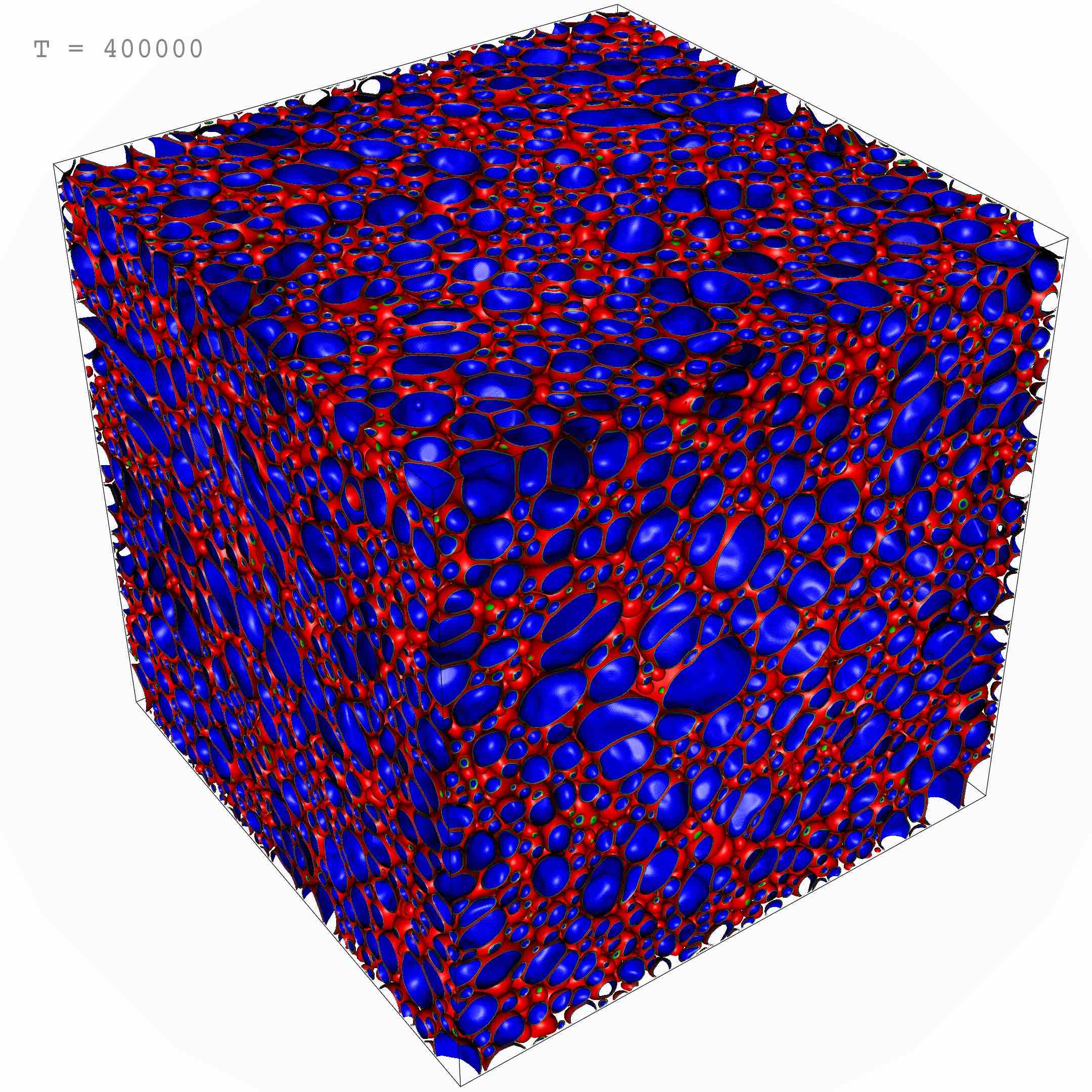}
  \label{figure:motion_h}
}
\subfigure[ $\phi=70\%$ $t=15 T_L$]{
  \includegraphics[width=0.17\textwidth]{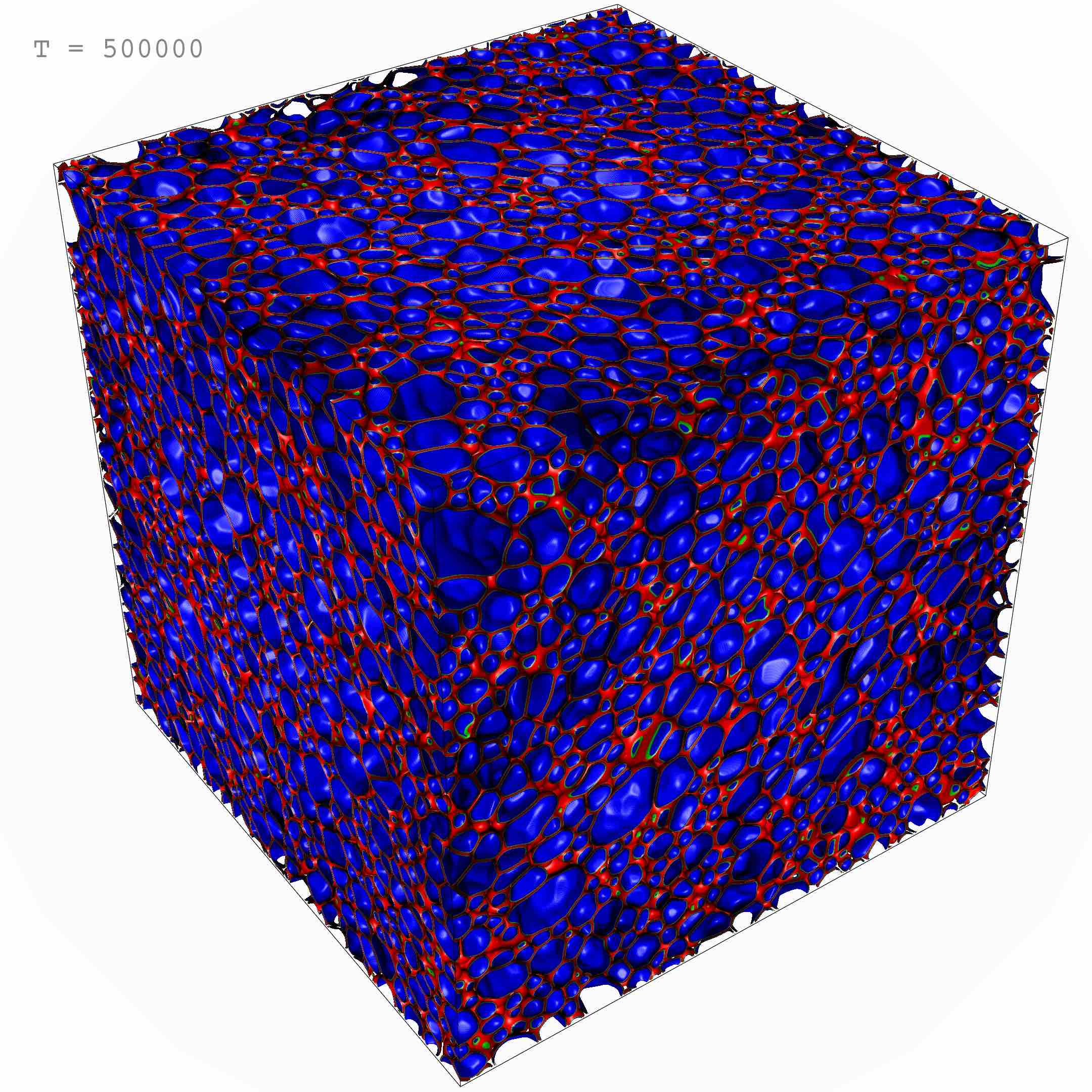}
  \label{figure:motion_i}
}
\subfigure[ $\phi=77\%$ $t=18 T_L$]{
  \includegraphics[width=0.17\textwidth]{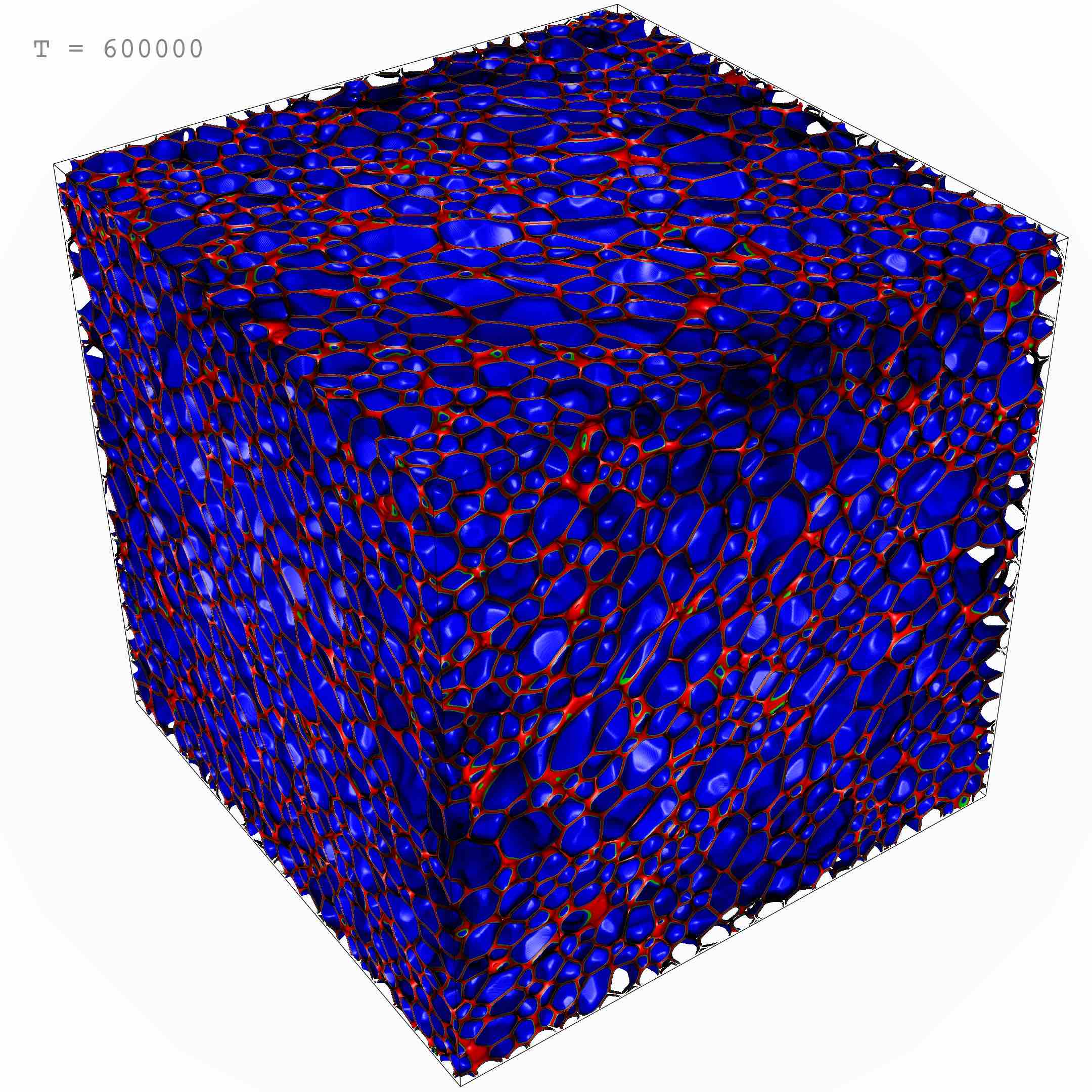}
  \label{figure:motion_l}
}
\caption{Graphical illustration of the emulsification process 
  via large-scale stirring and slow addition of the
  dispersed phase. 
  Snapshots of the interface (red/blue side corresponding to the continuous/dispersed phase) during stirring at various instants of time 
  (given in units of the large scale characteristic time $T_L = L/V_{\text{rms}}$, 
  where $V_{\text{rms}}$ is computed once the injection process is terminated, i.e. at the maximum volume fraction; 
  see also figure \ref{figure:n_droplets_norm} and caption therein). 
  Both time and the volume fraction of the dispersed phase $\phi$ grow from (a) to (j).
  Panel (a): the slightly deformed initially flat interface is still clearly visible,
  no droplets have formed yet. Panels (b)-(d): the process of
  fragmentation of the initial interface, leading to the
  production of a large number of small droplets,  can be appreciated (see e.g. panel
  (d)). Panels (e)-(j): $\phi$ further increases, droplets become smaller and the system
  becomes more and more densely packed. The simulation parameters are
  reported in Table \ref{table:runs} (run C).}
\label{figure:motion}
\end{figure}
In this article, starting with a fully phase-separated
system, we describe, by means of high-resolution numerical simulations,
the whole dynamics of the building up of a stabilized dense emulsion, of
which we monitor the evolution (in both the dense and semi-diluted
regimes) and measure the distribution of droplet sizes, during and
after stirring. In the early phase of the forced regime, a
power-law decay of the distribution with an exponent close to $-10/3$
is found, over a wide range of radii. Such scaling was found to be a
rather robust and universal feature of stirred multiphase systems,
from emulsions to the entrainment of air in the breaking of oceanic waves
\cite{garrett2000connection,deane2002scale,soligo2019breakage,mukherjee2019droplet}.
This $-10/3$ law was originally justified, for diluted systems,
resorting to dimensional analysis~\cite{garrett2000connection}.  Very
recently, it was derived, more rigorously, by means of energetic
arguments \cite{yu2020scale}.  In both cases, though, a
Kolmogorov-like turbulence phenomenology was invoked.  Observing the
same scaling in our jammed, non-turbulent, system challenges, then,
the basic theoretical understanding of flow stirred emulsions,
strongly pinpointing it as a still open problem. More generally
  we can accurately measure the time evolution of the droplet radii
  distribution function, and associated polydispersity, that we find
  to be an informative quantitative observable on the internal structure
  of the emulsions during the different phases of its dynamic
  evolution. We then rheologically characterize the emulsion,
highlighting the presence of a finite yield stress and, finally, we
discuss the possibility of probing the response of the system to
changes in the control parameters.

\section{Numerical setup and emulsification process}
\noindent We consider binary mixtures, where the two consituent fluids have identical 
physical properties, namely same density and viscosity.
We perform direct numerical simulations in tri-periodic computational domains of 
size $L^3$ (with different resolutions $L$); our study employs a numerical model based on a Lattice
Boltzmann method \cite{succi2001lattice} implementing a
nearest-neighbour lattice interaction of Shan-Chen type
\cite{shan1993lattice,shan1994simulation} which endows the system with
the nonideal character and gives rise to a surface tension.  A
next-to-nearest neighbours interaction is also introduced, which
promotes the emergence of a positive disjoining pressure within the
thin liquid films between close-to-contact interfaces, thus inhibiting
droplet coalescence
\cite{benzi2009mesoscopic,sbragaglia2012supramolecular}.  This
interaction provides, then, the required stabilising mechanism
(mimicking the role of surfactant in standard emulsions and of
nanoparticles in Pickering emulsions), that prevents the system from
full phase separation and keeps the emulsion in its (meta)stable
state.  The numerical model will not be discussed in further detail
here, since it has been extensively validated and applied to the study
on several physical problems, including rheology of confined foams
\cite{dollet2015twodimensional,scagliarini2015nonlocality}, plastic events
and ``avalanches'' in soft-glassy materials
\cite{benzi2015internal,benzi2016earthquake,pelusi2019avalanche},
thermal convection in emulsions \cite{pelusi2021rayleigh} and
microfluidics \cite{scagliarini2016fluidisation,pelusi2019roughness}.
Numerical values of the relevant parameters used in the numerical
simulations are reported in Table \ref{table:runs}.
\begin{table*}[htbp]
  \centering
  \begin{tabular}{c|c|c|c|c|c|c|c|c|c|c|c|c}
    $\text{RUN}_{\text{id}}$ & $L$ & $E_{\text{in}}$ & $V_{\text{rms}}$ & $T_L$ & $\tau_{\phi}/T_L$ & $\overline{N_D}$ & $R_{30}$  & $R_{10}$  & $R_{\sigma}$ & $Ca$ & $We$ & $Re$\\\hline
    \toprule
    A & 256 & $7.196\cdot{10^{-8}}$ & 0.0115 & 22261 &  35.9 & 112.7 & 30.1 & 26.7 & 9.50 & 0.109 & 0.227 & 123.47\\
    B & 512 & $6.294\cdot{10^{-8}}$ & 0.0178 & 28764 & 27.8 & 877.2 & 30.4 & 27.9 & 8.42 & 0.169 & 0.548 & 381.79\\
    C & 1024 & $6.597\cdot{10^{-8}}$ & 0.0312 & 32821 & 24.4 & 5707.1 & 32.6 & 28.4 & 11.03 & 0.296 & 1.809 & 1339.27\\
  \end{tabular}
  \caption{Main parameters and relevant observables for all simulations, labelled by a $\text{RUN}_{\text{id}}$, 
  presented in this paper: system size, $L$, (the grid spacing in all three directions is set to unity);
  energy injection rate, 
  $E_{\text{in}}=\overline{\mathbf{f} \cdot \mathbf{v}}$; 
  root mean square velocity, $V_{\text{rms}}=\sqrt{\overline{|\mathbf{v}|^2}}$; $T_L=L/V_{\text{rms}}$ is the large scale correlation time; 
  $\tau_{\phi}$ is the inverse injection rate;   average number of droplets, $\overline{N_D}$; 
  volume mean radius, $R_{30}=\sqrt[3]{\frac{3}{4\pi}\langle V_D\rangle}$
  ($V_D$ is the droplet volume); arithmetic mean radius, $R_{10}= \langle R \rangle$; standard deviation of droplet radii, 
  $R_{\sigma}$; capillary number, $Ca=\frac{\rho \nu {V_{\text{rms}}}}{\sigma}$; 
  Weber number, $We=\frac{\rho {R_{10}} {V_{\text{rms}}^2}}{\sigma}$; Reynolds number as 
  $Re=\frac{L V_{\text{rms}}}{\nu}$. $\rho=1.36$ is the total fluid density, $\nu=c^2_{s}(\tau_{\text{LB}}-0.5)=1/6$
  is the kinematic viscosity, equal for both pure fluids ($c_{s}=1/\sqrt{3}$ is the speed of sound and $\tau_{\text{LB}}=1$ 
  is the lattice Boltzmann relaxation time), and $\sigma=0.0238$ is the surface tension.
  The averages, $\overline{(\cdots)}$, are taken over time, in the interval $T_3$ 
  (where the final concentration of the emulsion is achieved but the emulsion is still stirred), whereas the averages 
  $\langle (\cdots) \rangle$ are meant taken over time and number of droplets.
  }
  \label{table:runs}
\end{table*}
A dense emulsion is characterized by a dispersed droplet phase with a
volume fraction $\phi$ in the order or larger than the one
corresponding to the random close packing of spheres ($\phi \sim 64\%$). Such
high volume fractions are achievable thanks to
the deformability of droplets and to the stabilization of interstitial
liquid films (as discussed above).
Making a dense emulsion out of two volumes of fluids, $V_1$ and $V_2$, separated by a flat interface (in a container of volume
$V=V_1+V_2$), is not an easy task.
When a stirring force is applied the interface is broken and droplets
are formed. However it is not possible to attain high droplet
concentration, of the majority phase, simply by stirring.  In fact, in
this way, the phase-inverted state is energetically more favorable,
whereby few droplets of the minority phase are dispersed in the
continuous majority phase \cite{vaessen1995applicability}.
Therefore, we follow a procedure that closely resembles some recipes
used to make mayonnaise at home. We start with a small volume
fraction, $\phi \sim 30\%$ of ``fluid one'' (say, e.g., oil), in a
larger volume fraction, $\sim 70\%$ of ``fluid two'' (water). The two
fluids are initially separated by a flat interface. At the beginning of the
simulation a large scale stirring is applied (as described in
\cite{biferale2011lattice}), with an amplitude strong enough to
break the initial interface and form droplets, but not too strong
 to destroy the emulsion (see simulation parameters in Table
\ref{table:runs}).  Under these conditions the forcing will deform (see
Figure \ref{figure:motion_a}) and break the flat interface in a
multitude of droplets, creating a low-volume fraction emulsion of oil
in water (see Figure
\ref{figure:motion_b}, \ref{figure:motion_c}, \ref{figure:motion_d}).
While the large scale forcing is active, we slowly add oil (fluid one)
and remove water (fluid two) in such a way to satisfy the constrain of total
volume conservation. In Figure \ref{figure:motion} the emulsions at
times corresponding to $\phi=30\%,31\%,32\%,33\%,35\%,42\%,50\%,60\%,70\%,77\%$ are visualized. 
As it can be seen the process of {\it adiabatically} adding oil,
while removing water, inflates the dispersed droplets already present
in the emulsion until these break in smaller droplets due to
hydrodynamics stresses induced by the large scale stirring. In this
way it is possible to constantly produce new droplets and pack them to prepare
an emulsion of high volume fractions. All the snapshots in Figure \ref{figure:motion}
are collected along the stirred run (when the forcing is applied). When the
forcing is switched off the emulsion relaxes to a resting state
and droplets achieve a more isotropic shape.\\ 
\begin{figure}[t!]
\centering
\subfigure[$512^3$ - $\phi=77\%$]{
  \includegraphics[width=0.2\textwidth]{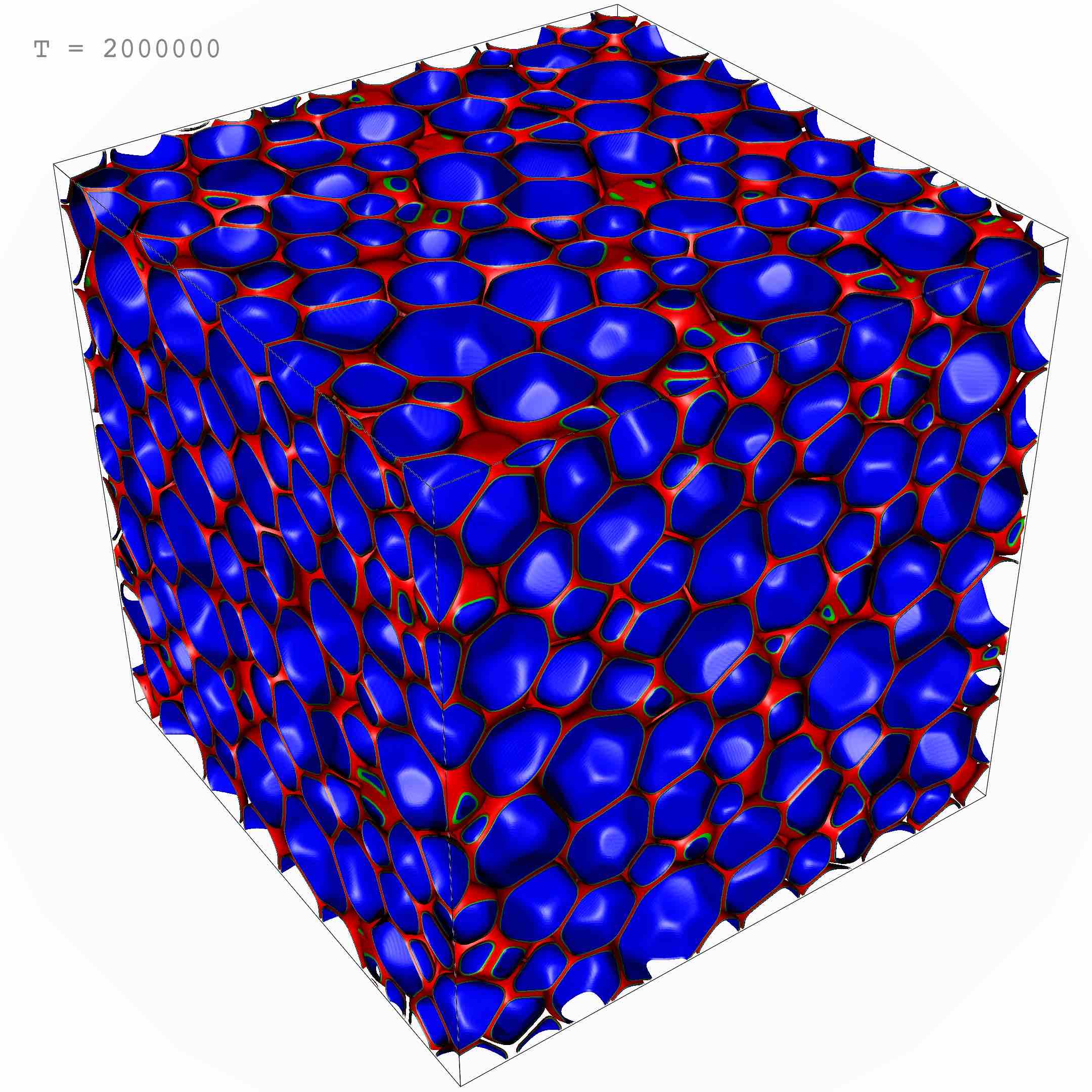}
  \label{figure:jammed_c}
}
\subfigure[$512^3$ - $\phi=28\% $]{
  \includegraphics[width=0.2\textwidth]{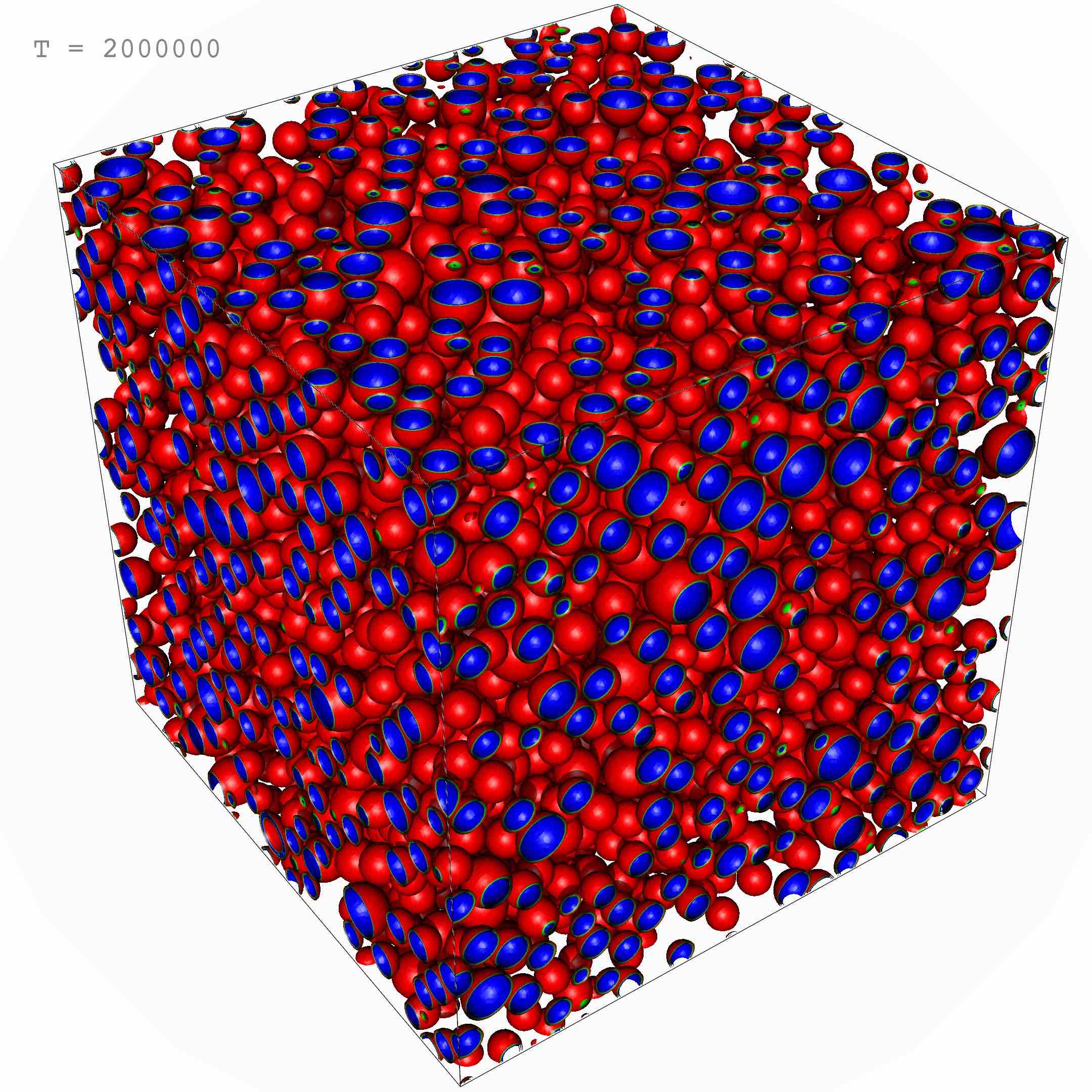}
  \label{figure:jammed_d}
}
\caption{Snapshot of the final interface field configuration from simulations with  
volume fraction $\phi=77\%$ (panel (a)) and $\phi=28\%$ (panel (b)).
As it can be observed, at the largest volume fraction the droplets 
are highly deformed, while at 
lower volume fraction they preserve their equilibrium spherical 
shape and their average size is smaller.} 
\label{figure:jammed}
\end{figure}
In Figure \ref{figure:jammed} one
can clearly appreciate the morphological difference between a high
concentrated (left, corresponding to $\phi=77\%$) and a low concentrated
emulsion (right, $\phi=28\%$):
while at low volume fraction the droplets maintain essentially their
spherical shape, at high volume fraction they are strongly deformed by
the dense packing and squeeze the continuous phase into the typical
network of foam-like structures.
\begin{figure}[!]
\centering
  \includegraphics[width=0.45\textwidth]{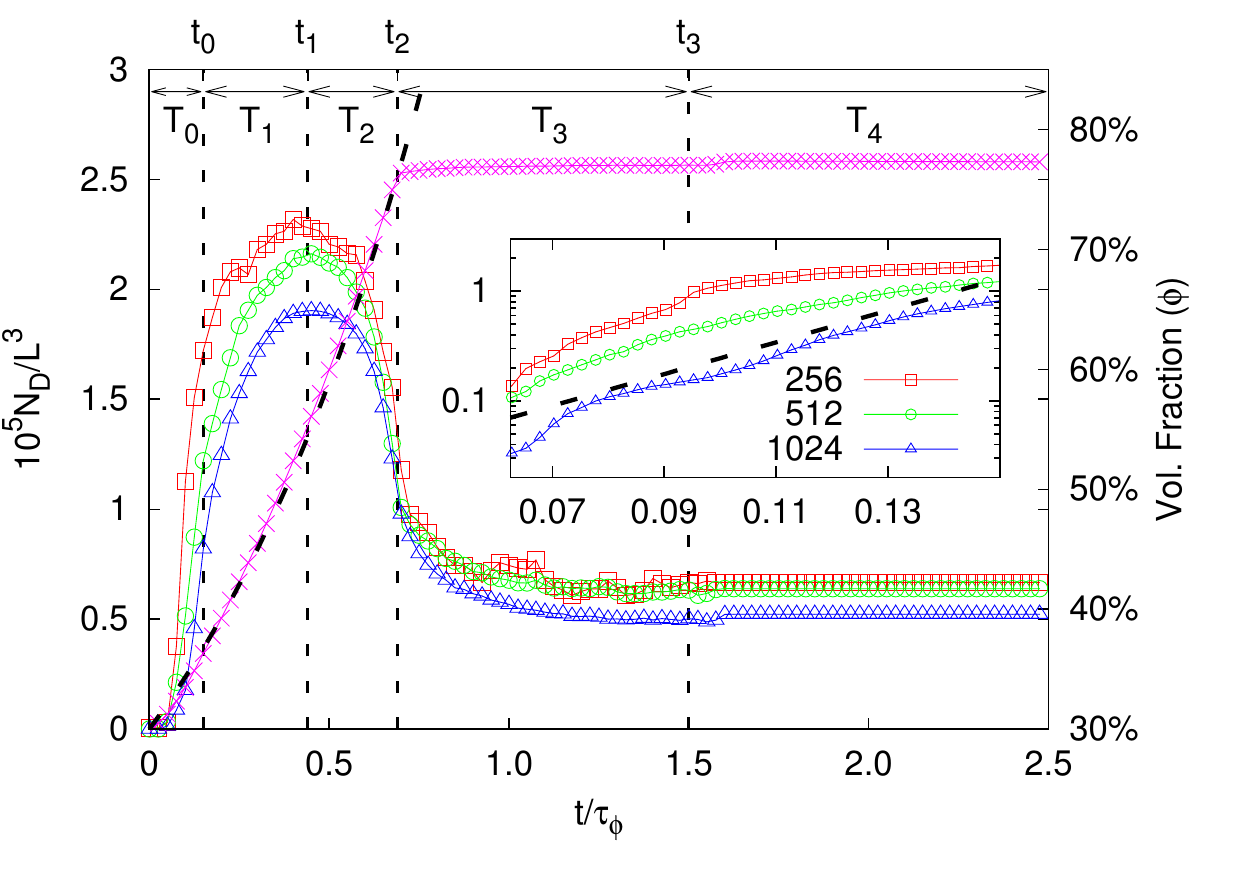}
  \caption{MAIN PANEL. Number density of droplets, $N_{D}/L^3$, as a function of time for different resolutions: $L=256$ (red squares, 
  \textcolor{red}{$\Box$}), $L=512$ (green circles, \textcolor{green}{$\circ$}) and 
  $L=1024$ (blue triangles, \textcolor{blue}{$\triangle$}). The volume fraction of the dispersed phase, $\phi(t)$, 
  as a function of time is also reported on the y2-axis (magenta crosses, \textcolor{magenta}{$\times$}), 
  together with an exponential fit in the 
  injection phase $T_0+T_1+T_3$ (dashed line). The time is given in units of $\tau_{\phi}$, the 
  inverse injection rate (i.e. such that the total mass of dispersed phase fulfills 
  $\dot{M}(t)=\tau_{\phi}^{-1} M(t)$).
  Further details on the parameter used in the simulations are provided in Table \ref{table:runs}.
  INSET: Zoom on the first time window ($T_0$) in logarithmic scale on the $y$-axis to highlight the initial exponential
  increase of the number of droplets (the dashed line represents the exponential fit, drawn as a guide to the eye).
}
\label{figure:n_droplets_norm}
\end{figure}

\begin{figure}[!]
\centering
  \includegraphics[width=0.37\textwidth]{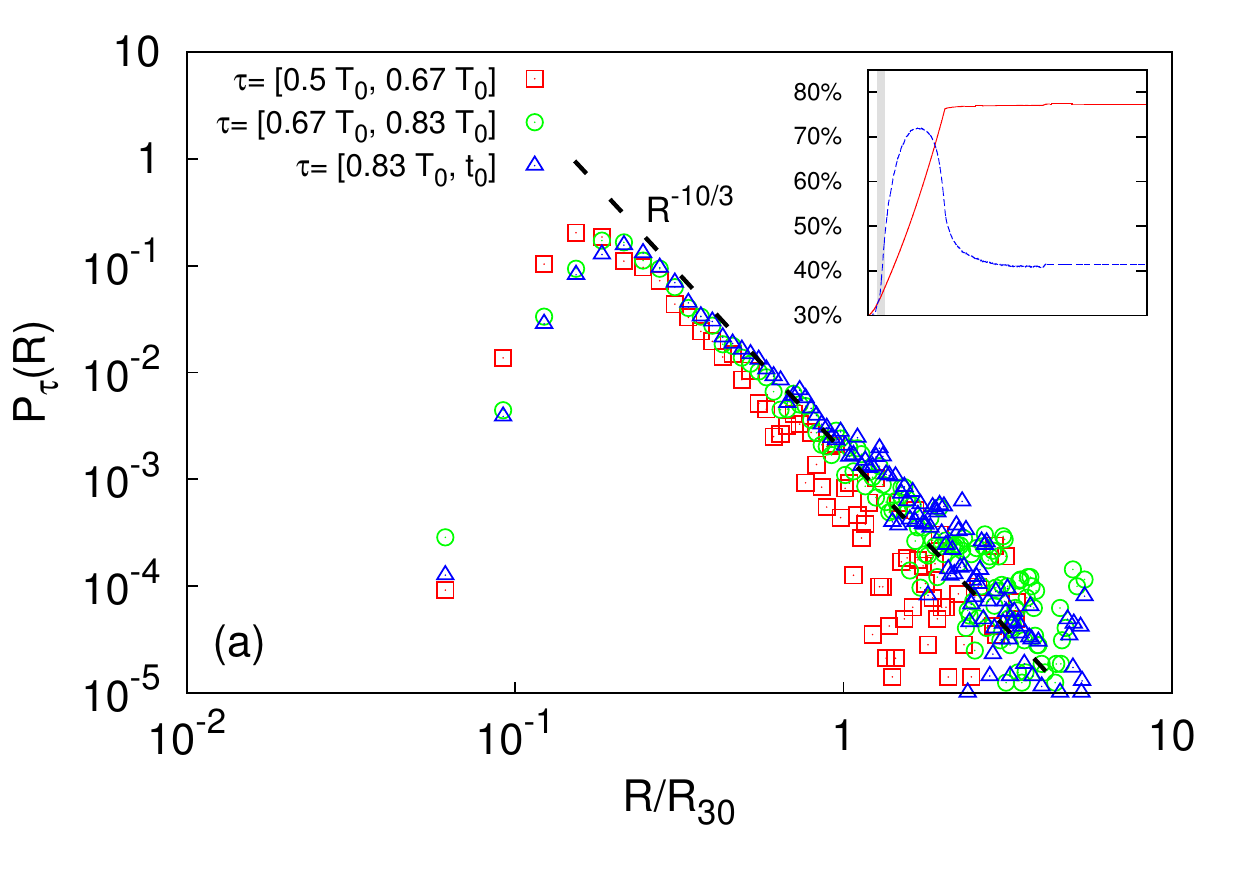}
  \label{figure:dsd_init_time_evolution_1}
  \includegraphics[width=0.37\textwidth]{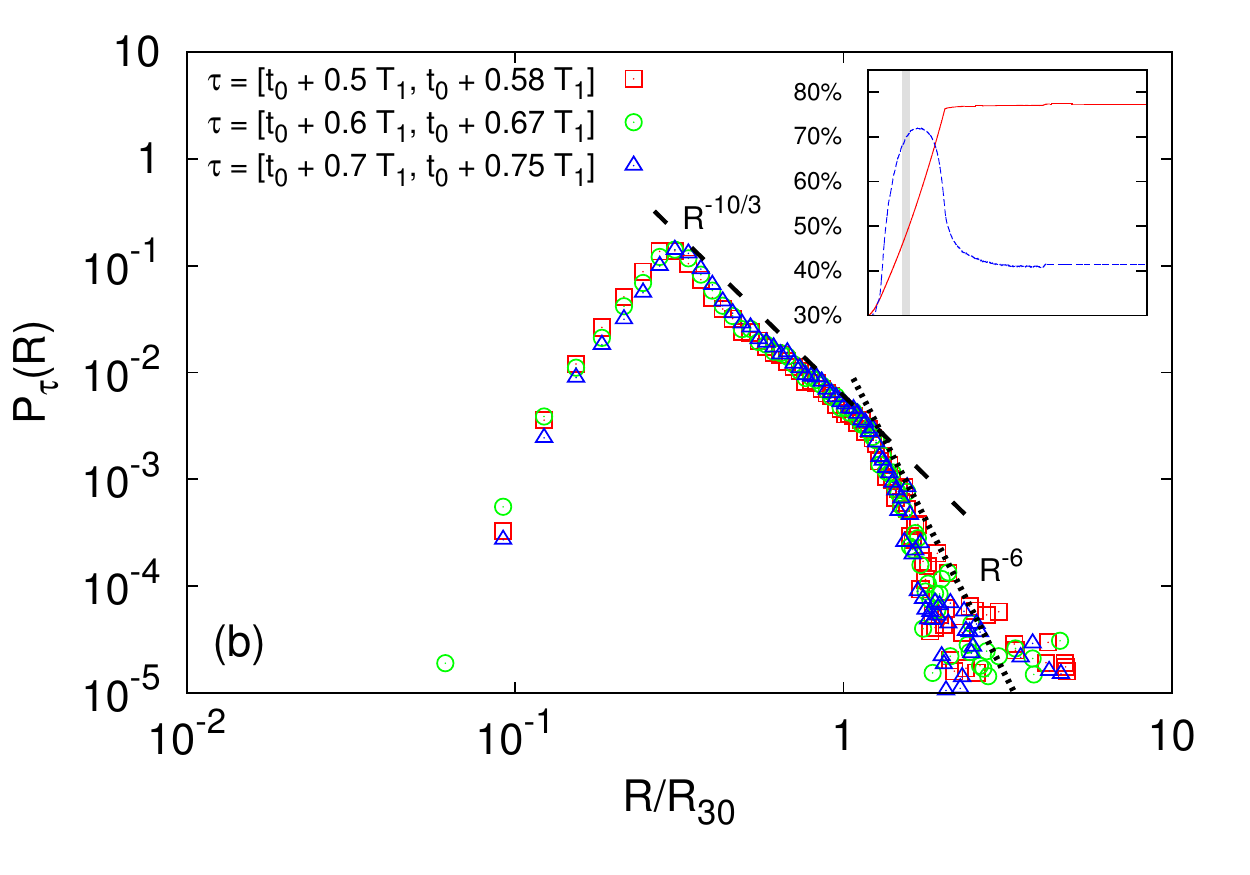}
  \label{figure:dsd_init_time_evolution_2}
  \includegraphics[width=0.37\textwidth]{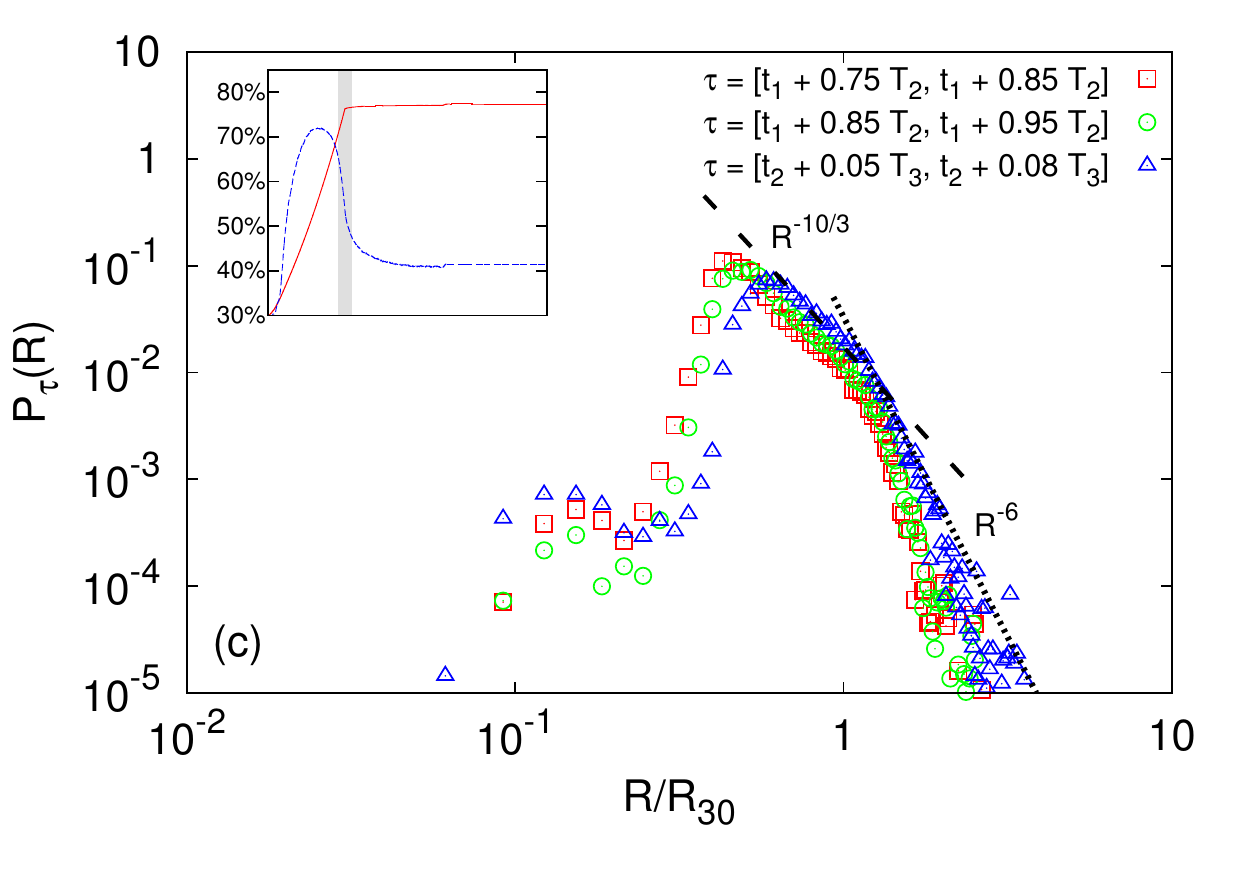}
  \label{figure:dsd_init_time_evolution_3}
  \includegraphics[width=0.37\textwidth]{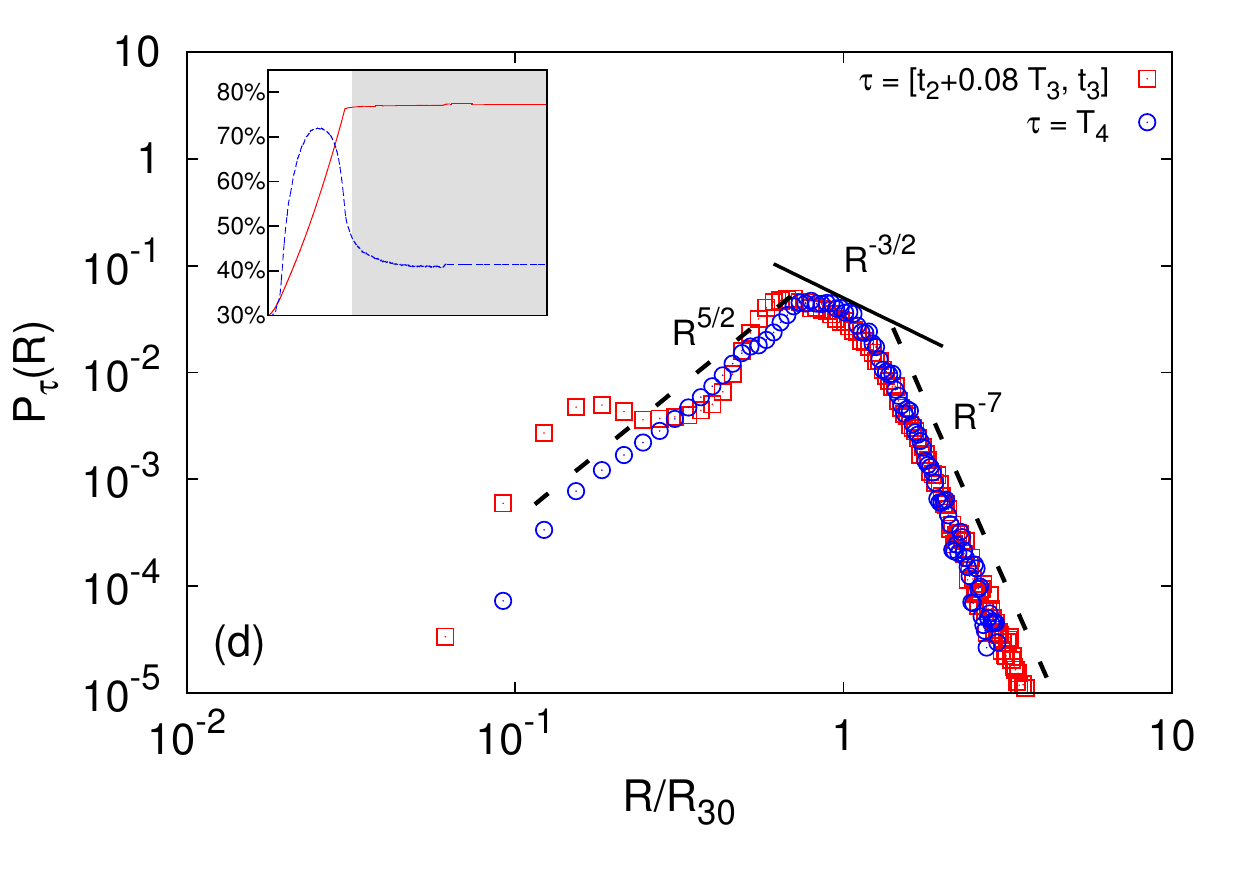}
  \label{figure:dsd_init_time_evolution_4}
\caption{PDFs $P_{\tau}(R)$ (from the simulation $C$, see Table \ref{table:runs}) of the effective droplet radii
(i.e. for each droplet, the radius of the equivalent sphere is measured) given in units of the mean volume radius $R_{30}$, 
computed in four different time windows, $\tau$.
  During the initial phase, $\tau \subset T_0$, the process of stirring-induced interface fragmentation dominates and 
  the PDF displays a peak at $R \approx 0.15 R_{30}$, followed by the typical slope $\sim R^{-10/3}$ (panel (a)).
  As the concentration of the dispersed phase increases, for $\tau \subset T_1$
  a steeper decay, $\sim R^{-6}$, develops for large sizes, the change of slop occurring at $R \approx R_{30}$; 
  in parallel, the peak is shifted towards higher $R$ (panel (b)).  
  Eventually, when the injection procedure is completed
  and the volume fraction of dispersed phase is $\approx 77\%$, the peak merges with the bend at $R\approx R_{30}$
  and a secondary peak emerges again at $R \approx 0.15 R_{30}$, giving $P_{\tau}(R)$ (for $\tau \subset T_3$) a bimodal shape, 
  evidence of the increasing presence of small droplets (panel (c)).
  This secondary maximum tends to vanish when the forcing is switched off (panel (d)).}
\label{figure:dsd_init_time_evolution}
\end{figure}

\section{Droplet size distribution}. 
\noindent The emulsion prepared with the procedure just described is
characterized by a microstructure which is not assembled {\it ad hoc},
as customary in the large majority of numerical studies of soft-glassy rheology,
but emerged, instead, {\it naturally} as the product
of the flow dynamics and hydrodynamic stresses. It is
therefore interesting to investigate the properties of the droplet size
distribution (DSD), during the initial transient phase as well for the final
resting dense emulsion. To our knowledge this is the first time that this investigation is done {\it during the emulsification process}.
In this work the strength of the large scale stirring force is not
enough to generate a fully developed turbulent flow, also due to the
fact that the emulsion, at increasing the volume fraction of the
dispersed phase, becomes more viscous. 
In Figure \ref{figure:n_droplets_norm} we show the total number of droplets 
vs time (shown in millions of LBE time steps for three different resolutions (with $L=256,512,1024$).
The magenta crosses show the time evolution of the volume fraction $\phi$ of the emulsion,
 see the right scale. At $t=0.6$ we stop increasing $\phi$ and keep constant at $\phi \sim 77\%$. 
Looking at the total number of droplets, $N_D$, we can distinguish $5$ different time windows in the system referred 
to as $T_i$, $i=0,\dots, 4$ and the corresponding ending times $t_k = \sum_{i=0}^k T_i$ (such that $T_k = t_k - t_{k-1}$). 
In the initial stage, $N_D$ grows quite rapidly during the time window $T_0$, 
it reaches a maximum in the time interval $[t_0, t_1]$ and then decreases in the interval $[t_1, t_2]$. 
At $t=t_2$ we stop increasing $\phi$ while retaining the large scale forcing. Finally  at $t=t_3$ we stop the large scale forcing.\\
In Figure \ref{figure:dsd_init_time_evolution} we 
show the probability distribution $P_{\tau}(R)$ of finding a droplet with radius $R$ during a time window $\tau$ 
(the four panels refer to different time windows). 
In panel (a), where $\tau \subset T_0$, a rather clear scaling $P_{\tau}(R) \sim R^{-10/3}$ is observed at relatively large $R$.
This result is remarkable because it closely overlaps with what
reported in the literature for breaking of oceanic waves and we
believe they cannot be supported here by an argument based on
turbulence inertial range scaling. 
Upon increasing the packing ratio, as shown in panels (b), for $\tau \subset T_1$, and (c), for 
$\tau \subset T_2$, the shape of the distribution changes: the range where $P_{\tau}(R) \sim R^{-10/3}$ is reduced, 
the peak is shifted towards larger values of $R$ and a new scaling behavior $P_{\tau}(R) \sim R^{-6}$ is observed, 
with the change of slope occurring at $R\approx R_{30}$. 
Eventually, during the time windows $T_3$ and $T_4$, where the packing ratio is kept constant, 
there is no evidence of the scaling $-10/3$ (see panel (d)).
It is interesting to observe that at the highest volume fraction, although for large values of $R$ 
the DSDs in the forced ($T_3$) and unforced ($T_4$) cases are basically indistinguishable, 
a significant difference emerges at small sizes:
under stirring, the presence of a second peak at $R \approx 0.15 R_{30}$ can be detected,
i.e. the $P_{T_3}(R)$ is bimodal. This secondary peak disappears when the
forcing is switched off. We attribute this peak to the large number of
breakup events for the high-volume fraction case. At further
increasing the volume fraction of the dispersed case, or at increasing
the intensity of the forcing, the emulsion rapidly breaks down via the
so-called catastrophic phase inversion.
Figure \ref{figure:dsd_init_time_evolution} conveys the important message that the DSD of 
an emulsion (hence its polydispersity) relies crucially, and in a highly non-trivial way, on 
the preparation protocol, namely on the stirring and injection rate and their time duration.
Consequently, our study suggests that, by leveraging on the parameters that characterize these processes, 
one may envisage to control the emulsion microstructure and hence the rheology.

\section{Emulsion rheology: yield stress and shear thinning}. 
\noindent We move now to the rheological characterization of the dense emulsions produced.
To this aim, we take the emulsion configuration obtained at time $t_4$
and we apply a force of the form $\mathbf{F}=(F_x(y),0,0)$, where
\begin{equation}\label{eq:fKolmo}
F_x(y) = F \sin\left(\frac{2\pi y}{L}\right),
\end{equation} 
and monitor the response of the system at changing the amplitude $F$.\\
We use the forcing in eqn. (\ref{eq:fKolmo}) because it is compliant with the 
periodic boundary conditions used during the emulsification process.
The shear stress induced by
(\ref{eq:fKolmo}) reads:
\begin{equation}\label{eq:stressK}
\sigma_{xy}(y) \equiv \sigma(y) = \int F_x(y')dy' = \frac{F L}{2\pi}
\cos \left(\frac{2\pi y}{L}\right).
\end{equation}
\begin{figure}[!]
\centering
\includegraphics[width=0.45\textwidth]{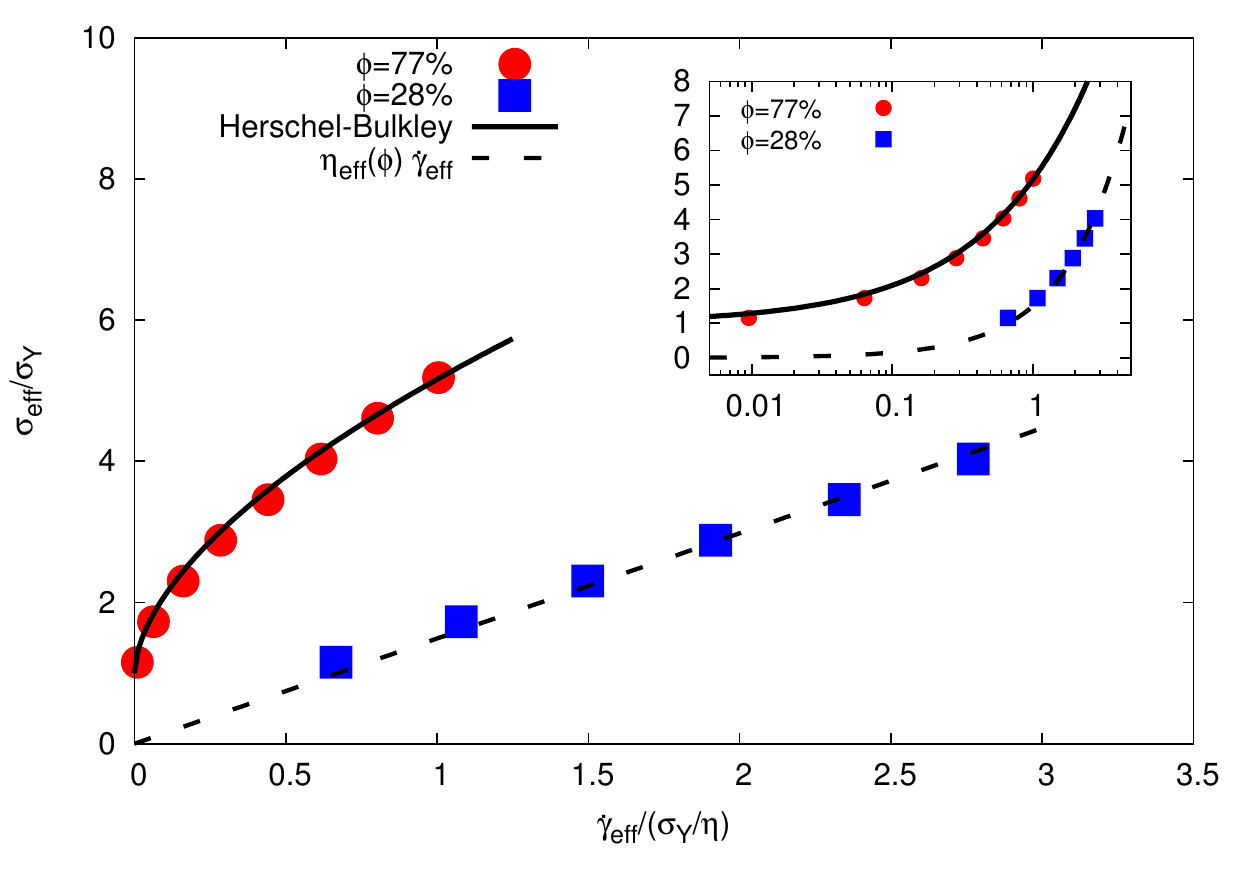} 
\caption{(Main panel) Flow curves for the emulsions with $\phi=77\%$ (red bullets) and $\phi=28\%$ 
  (blue squares), 
  showing effective shear stress, $\sigma_{\text{eff}}$ (normalized with the yield stress $\sigma_Y$), 
  vs effective shear rate, $\dot{\gamma}_{\text{eff}}$ (normalized with the yield stress divided by the dynamics viscosity 
  $\sigma_Y/\eta$), as defined in Eqs.(\ref{eq:effstress})-(\ref{eq:effshear}). The solid line indicates 
  a Herschel-Bulkley fit, $\sigma = \sigma_Y + K \dot{\gamma}^{\beta}$, 
  with $\sigma_Y = 2.5 \times 10^{-5}$ lbu, $K=0.02$ lbu,
  $\beta=0.58$, whereas the dashed line represents the Newtonian relation 
  $\sigma = \eta_{\text{eff}}\dot{\gamma}$, with an effective viscosity compliant with the Taylor's prediction for equiviscous, 
  low concentrated, emulsions, namely $\eta_{\text{eff}}(\phi)=\eta(1+(7/4)\phi)$. (Inset) Same as in 
  the main panel but in log-log scale.}
\label{fig:rheo}
\end{figure}
Following \cite{benzi2010herschel}, we can define an effective stress as
\begin{equation}\label{eq:effstress}
\sigma_{\text{eff}} = \langle \sigma^2 \rangle^{1/2} = \left(\frac{L}{2\pi}\right)\frac{F}{\sqrt{2}}
\end{equation}
while the corresponding effective shear rate $\dot \gamma_{\text{eff}}$ can be defined as
\begin{equation}\label{eq:effshear}
\dot{\gamma}_{\text{eff}} \equiv \frac{\langle \sigma(y)\dot{\gamma}(y)\rangle}{\langle \sigma^2(y)\rangle^{1/2}},
\end{equation}
where the average is meant to be taken over $y$, i.e. $\langle (\dots)\rangle = \frac{1}{L}\int_0^L (\dots)dy$.
From every simulation, with a certain forcing
amplitude $F$, we are able to extract a couple
$(\dot{\gamma}_{\text{eff}},\sigma_{\text{eff}})$. The resulting flow
curves are shown in Figure \ref{fig:rheo} for low and high volume
fractions; in both cases, obviously, $\sigma_{\text{eff}}$ grows with
$\dot{\gamma}_{\text{eff}}$, but while for $\phi=28\%$, as
$\dot{\gamma}_{\text{eff}} \rightarrow 0$, $\sigma_{\text{eff}}$
vanishes, for $\phi=77\%$ the stress tends to a finite ``yield''
value, $\sigma_Y$.  Remarkably, for the high volume fraction, the data
could be fitted very well with a relation of the Herschel-Bulkley
type, $\sigma = \sigma_Y + K \dot{\gamma}^{\beta}$, with $\sigma_Y
\approx 3 \times 10^{-5}$, $K \approx 0.02$ and $\beta \approx 0.6$ (denoting a shear-thinning character). 
In contrast, the less concentrated system shows a Newtonian character, 
$\sigma_{\text{eff}} = \eta_{\text{eff}}(\phi) \dot{\gamma}_{\text{eff}}$, though with an augmented (effective) viscosity 
that agrees with the Taylor expectation for diluted equiviscous emulsions, 
$\eta_{\text{eff}} \approx \eta(1+(7/4)\phi)$ (where $\eta=\rho \nu$ is the continuous phase dynamic viscosity) \cite{taylor1932viscosity}.
Before concluding, let us underline,
once more, that, in order to describe properly the rheology, 
it is crucial to have a realistic droplet size
distribution. As a step further, then, we will show, in future works,
how numerical simulations may help to correlate, quantitatively, the
parameters controlling the emulsion preparation (e.g., stirring force
amplitude, addition rate of the dispersed phase) to the those
characterising the emulsion rheological properties (e.g. the yield
stress).

\section{Conclusions}. 
\noindent We have demonstrated a numerical model and
approach that allows to study the formation and dynamics of dense
jammed emulsions with high space- and time-accuracy. The numerical
method describes two immiscible fluids with surface tension and
disjoining pressure, in order to stabilize droplets coalescence.
We focus on the production of a jammed emulsion by starting with a low
volume fraction, 30\%, of the dispersed phase and by continuous
stirring to fragment droplets in a chaotic flow. Slowly increasing the
dispersed phase we can achieve $~80\%$ volume fraction.
We measure the DSD and we show that soon after the breakup of the
initially flat interface we can clearly detect a distribution of large
droplet radii that seems in agreement with a $-10/3$ power-law
scaling. At later times, during the process of increasing the volume
fraction, the dynamics is dominated by droplets breakup event and the
DSD displays two distinct scaling behaviours with much steeper
exponents. At very large volume fraction, $~77\%$, the DSD displays
the emergence of a secondary peak around $1/10$ of the average droplet
radii. This peak is clearly associated with the dynamical flowing state, and ongoing
fragmentation processes, as it disappears once the large-scale stirring
is switched off.
While at the highest volume fraction achieved the system can still
flow, for the forcing intensities that we employ, we demonstrate that
this develops a finite yield stress. We show that such state is stable
under flow and that, switching off the stirring force, leads to a
jammed state with finite yield stress. Increasing the forcing
amplitude or the volume fraction of the dispersed phase leads to
catastrophic phase inversion, a topic that will be studied in a
forthcoming publication.  This numerical approach offer unique
perspectives to uncover the basic physics of dense emulsions, where
the mesoscopic dynamics of droplets is extremely difficult to be
studied via experimental techniques, and paves the way to the use of
simulations as a tool to guide a controlled emulsification, in order
to obtain emulsions with targeted structural and rheological
properties.

\bibliographystyle{unsrt}
\bibliography{paper.bib}

\end{document}